# Signature of superconductivity in pressurized La$_4$Ni$_3$O$_{10-x}$ single crystals grown at ambient pressure

Feiyu Li,‡ Yinqiao Hao,‡ Ning Guo,‡ Jian Zhang; Qiang Zheng,* Guangtao Liu,* and Junjie Zhang*

**ABSTRACT:** Nickelates have attracted enormous attention since the discovery of high-temperature superconductivity in La$_3$Ni$_2$O$_7$ under high pressure. However, whether superconducting nickelate single crystals can be prepared at ambient pressure remains elusive. Here we report signature of superconductivity in pressurized La$_4$Ni$_3$O$_{10-x}$ single crystals grown from potassium carbonate flux at ambient pressure. Single crystal X-ray diffraction and scanning transmission electron microscopy investigations revealed high-quality single crystals with perfect stacking of trilayers. Resistivity measurements indicate that the metal-to-metal transition observed at ambient pressure was suppressed under high pressure, and a sharp drop occurred at ~30 K at 77.9 GPa, consistent with superconductivity in pressurized La$_4$Ni$_3$O$_{10}$ single crystals grown by the floating zone method at an oxygen pressure of >18 bar. Our results not only provide an important path to prepare high-quality nickelate single crystals but also support superconductivity in nickelates under high pressure, promoting more systematic and in-depth research in this compelling field.

The mechanism of high-T$_c$ superconductivity in cuprates remains a big challenge.[1, 2] Design and synthesis of new high-T$_c$ superconductors in non-copper transition metal oxides are expected to provide new materials platforms to help understand the existing mystery.[3] Nickel resides next to copper in the periodic table, and layered nickelates were predicted to superconduct 25 years ago.[4] Similar to cuprates, charge/spin stripes, charge/spin density waves, and pseudogap have been found in layered nickelates.[5-7] Moreover, the square-planar trilayer metallic Pr$_4$Ni$_3$O$_8$, which exhibits key ingredients of superconducting cuprates including square lattice, low spin, and large orbital polarization, resembles 1/3 hole-doped cuprates.[8] Electron doped Pr$_4$Ni$_3$O$_8$ was predicted to superconduct.[9] Two years later, superconductivity in nickelates was discovered by Li et al. in thin films of infinite-layer Nd$_{0.8}$Sr$_{0.2}$NiO$_2$,[10] initiating the Nickel Age of Superconductivity.[6, 11-19] Superconductivity was reported in more infinite-layer nickelates including Pr$_{1-x}$Sr$_x$NiO$_2$,[20] La$_{1-x}$A$_x$NiO$_2$ (A=Sr, Ca),[21, 22] and Nd$_{1-x}$Eu$_x$NiO$_2$,[23] and quintuple-layer Nd$_6$Ni$_5$O$_{12}$,[24] all of them are thin films. Unfortunately, no superconductivity has been realized in bulk samples, including polycrystalline powders[25, 26] and single crystals,[27] limiting the research of nickelate superconductivity.

Recently, signature of high-T$_c$ superconductivity was reported by Sun et al. in Ruddlesden-Popper La$_3$Ni$_2$O$_7$ under pressure of >14 GPa,[28] significantly boosting nickelate superconductivity research.[28-72] Extensive efforts have been devoted;[28-72] however, some key fundamental questions remain elusive. These include:

(i) What is (are) the superconducting phase(s)? The growth of bilayer La$_3$Ni$_2$O$_7$ ("2222") single crystals requires high oxygen pressure and only a very narrow range of oxygen pressure (10-15 bar) works.[28, 68] In addition, the bilayer "2222" and hybrid "1313" (La$_2$NiO$_4$·La$_3$Ni$_2$O$_7$) polymorph competes during floating zone growth; thus, obtaining a pure phase is very challenging.[41, 60, 62] Both samples containing main phases of "2222" or "1313" were reported to show signature of superconductivity at ~80 K.[28, 62] Furthermore, other hybrid nickelates such as La$_2$NiO$_4$·La$_3$Ni$_2$O$_7$ ("1212"),[73] and short-range intergrowth exist in Ruddlesden-Popper nickelates,[62] making it more challenging to identify the origin of superconductivity.

(ii) Whether the observed high-T$_c$ superconductivity is bulk or filamentary in nature? Zhou et al. reported a superconducting volume fraction of ~1% in "2222" single crystals, consistent with filamentary superconductivity.[61] The absence of diamagnetic response in polycrystalline samples supports this scenario.[42] In contrast, Wang et al. reported an estimated superconducting volume fraction of ~97(10)% at 19 GPa in bilayer La$_2$PrNi$_2$O$_7$, indicating bulk superconductivity.[49]

(iii) Do other Ruddlesden-Popper nickelates superconduct under high pressure? The answer is yes.[30, 51, 63, 72] The trilayer La$_4$Ni$_3$O$_{10}$ has been reported to superconduct under high pressure with a superconducting volume fraction of >80% and is bulk superconductivity.[30]

(iv) Whether bulk high-T$_c$ superconductivity can be achieved at ambient pressure? Theoretical calculations predicted Tb$_3$Ni$_2$O$_7$[55] and Ac$_3$Ni$_2$O$_7$[45] as candidates. Up to date, none of them has been synthesized.

(v) Whether nickelate single crystals can be prepared at ambient pressure and also show superconductivity? For high-T$_c$ superconductivity in nickelate single-crystal samples, two "high pressure" conditions are required, i.e., high oxygen pressure is needed to prepare single crystals (10-15 bar for La$_3$Ni$_2$O$_7$ and 18-30 bar for La$_4$Ni$_3$O$_{10}$) and high pressure (>14 GPa) is required to induce superconductivity.[28, 30, 68] Previously we succeeded in growing La$_4$Ni$_3$O$_{10}$ single crystals at ambient pressure,[52] overcoming the first high-pressure barrier in nickelate superconductivity research. Whether the ambient-pressure grown single crystals superconduct under high pressure is a very important open question.

In this contribution, we report the signature of superconductivity in ambient-pressure flux grown single crystals of La$_4$Ni$_3$O$_{10-x}$. Our results reveal a new path to prepare high-quality superconducting nickelate single crystals.

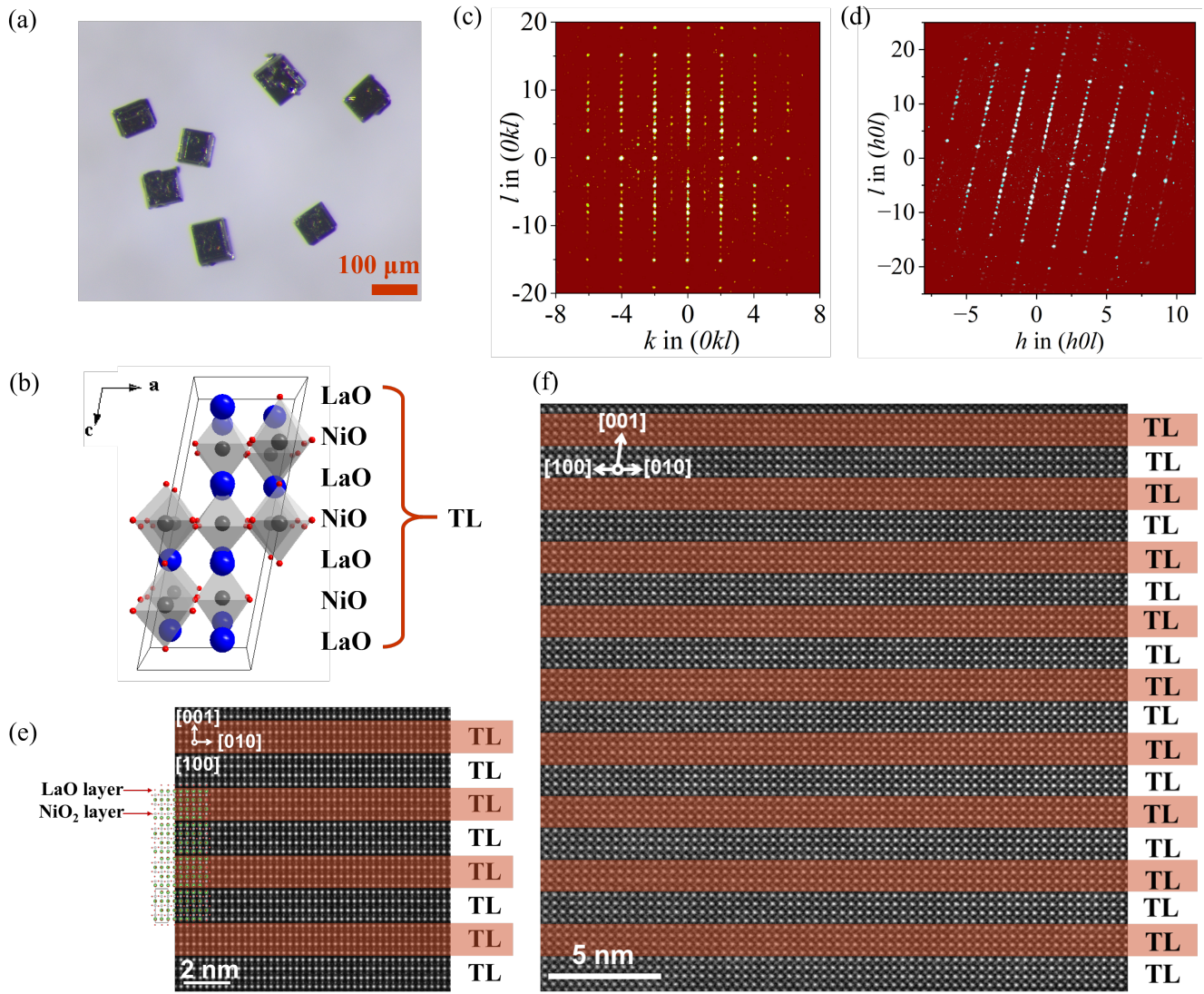

**Figure 1.** Crystallinity of $La_4Ni_3O_{10-x}$ single crystals grown at ambient pressure. (a) Photograph of typical as-grown $La_4Ni_3O_{10-x}$ single crystals. (b) Structural model of trilayer $La_4Ni_3O_{10}$ with LaO layers, $NiO_2$ layers and trilayer (TL) labeled. (c) Reconstructed ($0kl$) plane from single crystal X-ray diffraction data collected at 298 K. (d) Reconstructed ($h0l$) plane from single crystal X-ray diffraction data collected at 298 K. (e) A typical atomic-scale high-angle annular dark-field (HAADF)-STEM image along [100] projection with overlaid crystal structure and trilayer indicated. (f) A wide-range HAADF-STEM image along [110] with trilayer indicated.

$La_4Ni_3O_{10}$ crystals grown at ambient pressure were filtered first with a 400-mesh sieve to select single crystals with dimensions of >38 μm. Figure 1a shows a photograph of typical as-grown single crystals. Figure S1 shows the in-house X-ray powder diffraction data and Rietveld refinement of pulverized as-grown single crystals using $P2_1/a$. The refinement converged to $R_{wp}$ = 3.87%, $R_{exp}$ = 2.98%, and GOF = 1.30. The obtained lattice parameters for $La_4Ni_3O_{10}$ are $a$= 5.4185(1) Å, $b$= 5.4678(1) Å, $c$=14.2296(1) Å, and $\beta$= 100.7245(2)°, consistent with previous reports.[52, 68] It is worth noting that even filtered, the sample still contains ~4 wt % $La_3Ni_2O_7$, which is from a small amount of $La_3Ni_2O_7$ single crystals. Similar phenomena have been found in the flux growth of "1212" single crystals.[73]

We then characterized the crystalline quality of as-grown $La_4Ni_3O_{10}$ single crystals using in-house single crystal X-ray diffraction and scanning transmission electron microscopy (STEM), which assess the average structure and local structure, respectively. Figure 1b shows the structural model of $La_4Ni_3O_{10}$ in the $ac$ plane (space group $P2_1/a$). The structure is quasi-two-dimensional, consisting of trilayer perovskite layers alternating with rocksalt layers perpendicular to the $ab$ plane. Single crystal X-ray diffraction confirmed that the as-grown $La_4Ni_3O_{10}$ single crystals are well crystallized. Figures 1c, d show the reconstructed ($0kl$) and ($h0l$) planes from a total of 2971 frames from single-crystal X-ray diffraction data collected at 298 K. The sharp diffraction peaks and neat distribution of reflections demonstrate high crystallinity of $La_4Ni_3O_{10-x}$ single crystals.

Real-space imaging of the structure of $La_4Ni_3O_{10-x}$ single crystals was performed using STEM. Figure 1e shows a typical atomic-scale high-angle annular dark-field (HAADF)-STEM image along [100] projection with trilayer indicated, and Figure 1f shows a wide-range HAADF-STEM image along [110] with trilayer indicated. As can be seen clearly, $La_4Ni_3O_{10-x}$ single crystals grown at ambient pressure maintain a perfect stacking of trilayer perovskite layers alternating with rocksalt layers in the long-range order (see Figure S2-S5 for more HAADF-STEM images in SI). No hybrid Ruddlesden-Popper such as

"1313" or "1212", or short-range intergrowth were found.[41, 60, 62, 73]

Resistivity and magnetic susceptibility were measured at ambient pressure under various magnetic fields. The as-grown single crystals with dimensions of ~100 μm are too small to put four leads on so we measured resistivity on polycrystalline pellets using the conventional four-probe method by crushing single crystals. The pellets were sintered at 900 ºC in the air for 10 h to increase density. Powder X-ray diffraction confirmed no structural change after sintering (Figure S6). Figure 2a shows the resistivity as a function of temperature. A clear metal-to-metal transition at ~135 K is observed, consistent with previous reports.[30, 52] Magnetic susceptibility was measured on the same sample, as shown in Figure 2b. Magnetic susceptibility decrease as a function of temperature and shows an anomaly at ~135 K, consistent with the spin-density-wave transition.[7] No obvious magnetic field dependence of the transition was found in both resistivity and magnetic susceptibility measurements (Figure S7-S8).

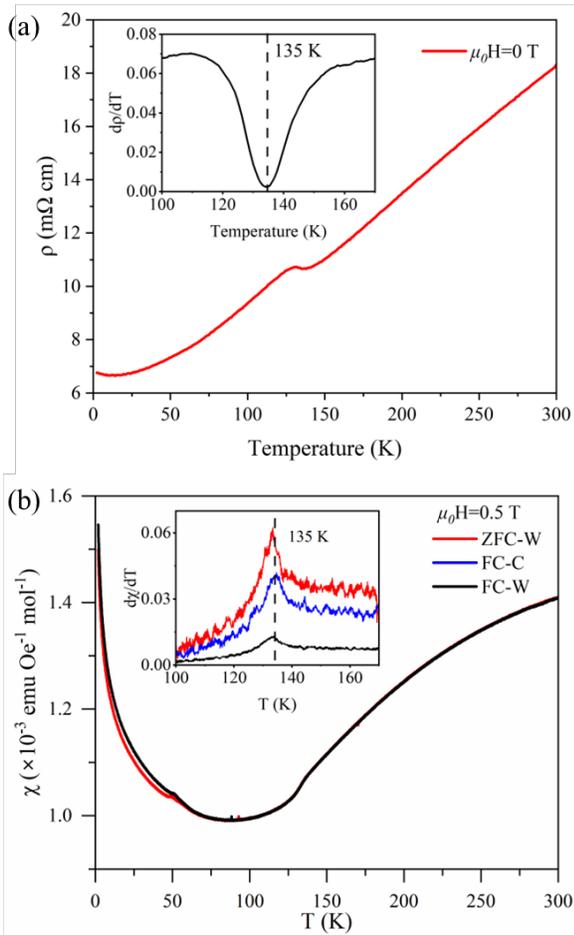

**Figure 2**. Resistivity (a) and magnetic susceptibility (b) of $La_4Ni_3O_{10-x}$ at ambient pressure. ZFC-W: zero-field cooling with data collected on warming; FC-C: field cooling and data collected on cooling; FC-W: field cooling and data collected on warming.

Resistivity measurements were carried out on single crystals under high pressure to explore superconductivity. To maintain consistency with the physical property measurements at ambient pressure, the as-grown $La_4Ni_3O_{10-x}$ single crystals were treated with identical conditions. The annealed single crystals were then subjected to resistance measurements under various pressures. Figure 3a shows the temperature-dependent resistances of sample #1 under pressure ranging from 4.8 to 77.9 GPa. At 4.8 GPa, the known metal-to-metal transition is clearly seen, and the transition temperature ($T_{MMT}$) is ~107.5 K, much lower than that at ambient pressure. As the pressure increases, the resistance decreases significantly, and the metal-to-metal transition gradually disappears, accompanied by a sharp drop in $T_{MMT}$ (Figure 3a). When the pressure increases to 13.6 GPa, the metal-to-metal transition is barely observed, and a signature of superconductivity with a $T_c$ (onset) of ~7 K emerges. With further compression, the $T_c$ (onset) gradually increases, reaching the maximum of 29.6 K at 77.9 GPa in this study. Although zero resistance is not directly observed due to the complexity of nickelate superconductors, as originally reported in $La_3Ni_2O_7$,[28] transport experiments under magnetic fields could provide key evidence for verifying their superconductivity. Two mechanisms, namely the Pauli paramagnetic effect and the orbital diamagnetic effect, break the Cooper pairs of the superconductor, resulting in a decrease in $T_c$ in the presence of an external magnetic field. Figure 3b shows the field dependence of the resistance of sample #1 at 57.9 GPa. It can be seen that $T_c$ is significantly suppressed with an increasing magnetic field up to 9 T. By utilizing the criteria of 90% of the normal state resistance values, we extrapolated the value of upper critical magnetic field $\mu_0H_{c2}(T)$ at 59.7 GPa and zero temperature to be 24 T (Figure 3c), and calculated the t zero-temperature coherence length $\xi(0)$ to be 37 Å, using the Ginzburg-Landau (GL) model.[74, 75] Given the high upper critical magnetic field, $La_4Ni_3O_{10-x}$ can be classified as a type-II superconductor. To repeat the conductive behavior of the flux-grown $La_4Ni_3O_{10-x}$ sample, the transport experiments of Sample #2 were conducted under pressures ranging from 2.7 to 21.8 GPa, and the temperature-dependent resistance is shown in Figure 3d. The metal-to-metal transition and superconducting signature were successfully reproduced with similar $T_{MMT}$ and $T_c$ (onset), further confirming the reliability of our results.

To investigate the relationship between the emergence of superconductivity of $La_4Ni_3O_{10-x}$ and its crystal structure under high pressure, in-situ synchrotron X-ray powder diffraction measurements were performed up to 22.5 GPa at ambient temperature (Figure S9). $La_4Ni_3O_{10-x}$ undergoes a pressure-induced structural transition between 2.3 and 9.3 GPa, as indicated by the clear variation in XRD peaks in the 2θ range of 12-19.9° (Figure 3e). The pressure of this structural phase transition is lower compared with the value in the transport measurements at which the metal-to-metal transition disappears and superconductivity emerges, inferring that the pressure-induced symmetry change is prerequisite to the emergence of superconductivity in $La_4Ni_3O_{10-x}$. Our observation is consistent with Wang et al.'s report on $La_2PrNi_2O_7$.[49] The crystal structure of $La_4Ni_3O_{10-x}$ under high pressure has been reported to be *I4/mmm* by Zhu et al.[30] and Li et al.[76] We thus refined our data using *I4/mmm*, and the results are reasonable (Figure S10 and Table S1). We also performed Rietveld refinements on our data using *Fmmm* and *Cmce* space groups considering that the peaks are relatively broad. It turns out that the refinements are also reasonable, and the lengths of the two short axes do not merge, as expected for *I4/mmm*. Therefore, within our resolution of measurements, it is rather difficult to determine the crystal structure under high pressure, and high-resolution X-ray single-crystal diffraction is required to address this structural puzzle.

We summarized the high-pressure resistance data and XRD data in Figure 3f to show the phase diagram of $La_4Ni_3O_{10-x}$. At ambient pressure, $La_4Ni_3O_{10-x}$ undergoes a metal-to-metal transition originating from charge and spin density wave order.[7] This density wave order is rapidly suppressed with increasing of pressure. Before the pressure reaches 9.3 GPa, a structural transition from monclinic $P2_1/a$ to high symmetry occurs. Evidence of superconductivity shows up after the structural transition, with a maximum $T_c$ of ~30 K at 77.9 GPa. Our results on ambient-pressure grown single crystals are consistent with those of Zhu et al.[30] on $La_4Ni_3O_{10}$ single crystals prepared by floating zone furnace at an oxygen pressure of 18-22 bar.

The resistance data at different pressures show that our samples have good reproducibility. However, our measurements have not reached zero resistance. There are two possible explanations for this. First, it may be related to the pressure transfer medium. As previously reported, Zhang et al.[32] observed zero resistance of $La_3Ni_2O_7$ while Sun et al.[28] did not, and Zhu et al.[30] observed zero resistance in $La_4Ni_3O_{10}$ while Zhang et al. did not.[28] We are trying helium and others as pressure mediums to see if zero resistance can be achieved. Second, it is possible that our samples are oxygen deficient, i.e., the oxygen content is less than 10. This is due to the fact that our growth method is carried out in a flux under ambient pressure. Although the samples were annealed in air, they may still have an oxygen deficiency compared with those grown by the high oxygen pressure floating zone method. Tuning oxygen content in flux-grown $La_4Ni_3O_{10-x}$ single crystals and exploration of its effect on superconductivity are in progress.

In conclusion, we performed high-pressure electrical resistance measurements on high-quality $La_4Ni_3O_{10-x}$ single crystals grown at ambient pressure. Single crystal X-ray diffraction data and STEM data demonstrate that our samples are long-range ordered single crystals with perfect stacking of trilayers. Electrical resistance measurements under high pressure show suppression of charge/spin density wave order and signs of emergence of superconductivity. The $La_4Ni_3O_{10}$ crystals obtained without high oxygen pressure show the same phenomena as the $La_4Ni_3O_{10}$ single crystals grown by the high-pressure floating zone method. We have overcome the first "high pressure" condition to synthesize superconducting nickelate single crystals. This work sheds important light on advancing the understanding of high-temperature superconductivity in nickelates.

Note: During the preparation of this manuscript, we became aware of a paper by Shi et al (arXiv: 2501.12647) who reported the synthesis and characterization of monoclinic and tetragonal $La_4Ni_3O_{10}$ microcrystals using flux method.

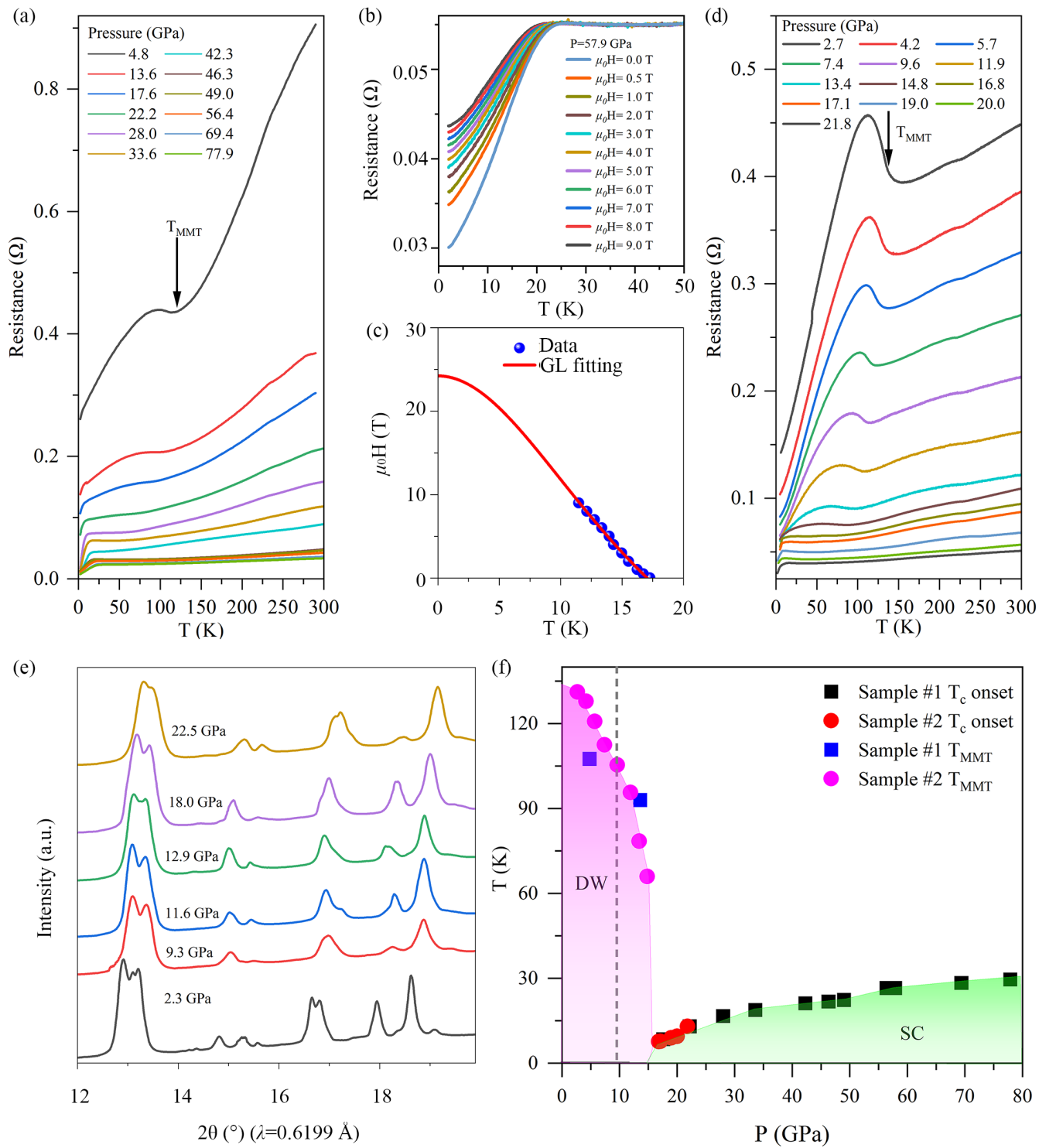

**Figure 3.** Electrical resistance and X-ray diffraction of $La_4Ni_3O_{10-x}$ under high pressure. (a) Temperature-dependent resistance of Sample #1 at 4.8-77.9 GPa. (b) Field dependence of electrical resistance at 57.9 GPa for sample #1. (c) The Ginzburg-Landau (GL) fitting of the upper critical field at 57.9 GPa (Ginzburg-Landau formula $H_{c2}(T)= H_{c2}(0)[(1-t)^2/(1+t^2)]$, where $t= T/T_c$). (d) Temperature-dependent resistances of Sample #2 at 2.7-21.8 GPa. (e) X-ray powder diffraction data of pulverized $La_4Ni_3O_{10-x}$ single crystals in the 2θ range of 12-19.9° ($\lambda$=0.6199 Å) at room temperature under various pressure. (f) Phase diagram of $La_4Ni_3O_{10-x}$ under high pressure. The black solid square and the red solid circle represent the $T_c$ onset of sample #1 and sample #2, respectively. $T_c$ onset is defined as the temperature at which the resistance deviates from its linear dependence at high temperature. The blue solid square and the magenta solid circle indicate the metal-to-metal transition temperature of sample #1 and sample #2, i.e., $T_{MMT}$. The dash line shows the pressure of 9.3 GPa around which the structural transition from $P2_1/a$ to high symmetry occurs.

## ASSOCIATED CONTENT

**Supporting Information**.

The Supporting Information is available free of charge at http://pubs.acs.org.

**Experimental details** for single crystal growth, powder X-ray diffraction, single crystal X-ray diffraction, resistivity and magnetic susceptibility at ambient pressure, scanning transmission electron microscopy, powder XRD under high pressure and resistance measurements under high pressure. **Figure S1.** In-house X-ray powder diffraction and rietveld refinement of pulverized as-grown single crystals. **Figure S2.** Real-space imaging of $La_4Ni_3O_{10-x}$ along [100] and intensity profile. **Figure S3.** Real-space imaging of $La_4Ni_3O_{10-x}$ along [100]. **Figure S4.** Real-space imaging of $La_4Ni_3O_{10-x}$ along [110] and intensity profile. **Figure S5.** Real-space imaging of $La_4Ni_3O_{10-x}$ along [110] and intensity profile. **Figure S6.** In-house X-ray powder diffraction and rietveld refinement of pulverized single crystals after sintering in air. **Figure S7.** Magnetic susceptibility of $La_4Ni_3O_{10-x}$ under various magnetic fields at ambient pressure. **Figure S8.** Resistivity of $La_4Ni_3O_{10-x}$ under various magnetic fields at ambient pressure. **Figure S9.** PXRD data of pulverized $La_4Ni_3O_{10-x}$ single crystals under high pressure at room temperature. **Figure S10.** Rietveld refinements on powder XRD data under high pressure. **Table S1.** Summary of structural details using various models under high pressure.

## AUTHOR INFORMATION

### Corresponding Author

**Junjie Zhang** – State Key Laboratory of Crystal Materials and Institute of Crystal Materials, Shandong University, Jinan, Shandong 250100, China; orcid.org/0000-0002-5561-1330; Email: junjie@sdu.edu.cn

**Guangtao Liu** – Key Laboratory of Material Simulation Methods and Software of Ministry of Education, College of Physics, Jilin University, Changchun 130012, China; orcid.org/0000-0003-2332-115X; Email: liuguangtao@jlu.edu.cn

**Qiang Zheng** – CAS Key Laboratory of Standardization and Measurement for Nanotechnology, CAS Center for Excellence in Nanoscience, National Center for Nanoscience and Technology, Beijing 100190, China; orcid.org/0000-0002-9279-7779; Email: zhengq@nanoctr.cn

### Authors

**Feiyu Li** – State Key Laboratory of Crystal Materials and Institute of Crystal Materials, Shandong University, Jinan, Shandong 250100, China; orcid.org/0009-0009-1254-3583.

**Yinqiao Hao** – State Key Laboratory of Superhard Materials and International Center of Computational Method & Software, College of Physics, Jilin University, 130012, Changchun, China; orcid.org/0009-0002-8005-1715.

**Ning Guo** – CAS Key Laboratory of Standardization and Measurement for Nanotechnology, CAS Center for Excellence in Nanoscience, National Center for Nanoscience and Technology, Beijing 100190, China.

**Jian Zhang** – State Key Laboratory of Crystal Materials and Institute of Crystal Materials, Shandong University, Jinan, Shandong 250100, China; orcid.org/0000-0002-1725-0466.

### Author Contributions

[‡]F.L. Y. H. and N.G. contributed equally.

### Notes

The authors declare no competing financial interest.

## ACKNOWLEDGMENT

J.Z. is grateful to Prof. Xutang Tao from Shandong University for providing valuable support. J.Z. thanks Dr. Yu-Sheng Chen and Dr. Tieyan Chang from University of Chicago for stimulating discussions. J.Z. thanks Prof. Jinguang Cheng from Institute of Physics Chinese Academy of Sciences for early collaboration on high pressure measurement. Work at Shandong University was supported by the National Natural Science Foundation of China (Grant No. 12074219). G.L. was supported by the National Natural Science Foundation of China (Grant No. 12074139). Q.Z. was supported by the National Key Basic Research Program of China under Grant No. 2021YFA1202801. J.Z. also acknowledges the support by the TaiShan Scholars Project of Shandong Province (Grant No. tsqn201909031) and the QiLu Young Scholars Program of Shandong University.

## REFERENCES

(1) Keimer, B.; Kivelson, S. A.; Norman, M. R.; Uchida, S.; Zaanen, J. From quantum matter to high-temperature superconductivity in copper oxides. *Nature* **2015**, *518* (7538), 179-186.
(2) Sanders, S. *125 Questions: Exploration and Discovery*; Science, 2021.
(3) Norman, M. R. Materials design for new superconductors. *Rep. Prog. Phys.* **2016**, *79* (7), 074502.
(4) Anisimov, V. I.; Bukhvalov, D.; Rice, T. M. Electronic structure of possible nickelate analogs to the cuprates. *Phys. Rev. B* **1999**, *59* (12), 7901-7906.
(5) Ishizaka, K.; Taguchi, Y.; Kajimoto, R.; Yoshizawa, H.; Tokura, Y. Charge ordering and charge dynamics in $Nd_{2-x}Sr_xNiO_4$ ($0.33 \leq x \leq 0.7$). *Phys. Rev. B* **2003**, *67* (18), 184418.
(6) Zhang, J.; Tao, X. Review on quasi-2D square planar nickelates. *CrystEngComm* **2021**, *23* (18), 3249-3264.
(7) Zhang, J.; Phelan, D.; Botana, A. S.; Chen, Y.-S.; Zheng, H.; Krogstad, M.; Wang, S. G.; Qiu, Y.; Rodriguez-Rivera, J. A.; Osborn, R.; Rosenkranz, S.; Norman, M. R.; Mitchell, J. F. Intertwined density waves in a metallic nickelate. *Nat. Commun.* **2020**, *11* (1), 6003.
(8) Zhang, J.; Botana, A. S.; Freeland, J. W.; Phelan, D.; Zheng, H.; Pardo, V.; Norman, M. R.; Mitchell, J. F. Large orbital polarization in a metallic square-planar nickelate. *Nat. Phys.* **2017**, *13* (9), 864-869.
(9) Botana, A. S.; Pardo, V.; Norman, M. R. Electron doped layered nickelates: spanning the phase diagram of the cuprates. *Phys. Rev. Mater.* **2017**, *1* (2), 021801.
(10) Li, D.; Lee, K.; Wang, B. Y.; Osada, M.; Crossley, S.; Lee, H. R.; Cui, Y.; Hikita, Y.; Hwang, H. Y. Superconductivity in an infinite-layer nickelate. *Nature* **2019**, *572* (7771), 624-627.
(11) Botana, A. S.; Lee, K.-W.; Norman, M. R.; Pardo, V.; Pickett, W. E. Low Valence Nickelates: Launching the Nickel Age of Superconductivity. *Front. Phys.* **2022**, *9*, 813532.
(12) Gu, Q.; Wen, H.-H. Superconductivity in nickel-based 112 systems. *The Innovation* **2022**, *3* (1), 100202.
(13) Hepting, M.; Dean, M. P. M.; Lee, W.-S. Soft X-Ray Spectroscopy of Low-Valence Nickelates. *Front. Phys.* **2021**, *9* (777), 808683.
(14) Ji, Y.; Liu, J.; Li, L.; Liao, Z. Superconductivity in infinite layer nickelates. *J. Appl. Phys.* **2021**, *130* (6), 060901.
(15) Li, D. The discovery and research progress of the nickelate superconductors. *Sci. Sin-Phys. Mech. Astron.* **2021**, *51* (4), 047405.
(16) Nomura, Y.; Arita, R. Superconductivity in infinite-layer nickelates. *Rep. Prog. Phys.* **2022**, *85* (5), 052501.
(17) Wang, B. Y.; Lee, K.; Goodge, B. H. Experimental Progress in Superconducting Nickelates. *Annu. Rev. Condens. Matter Phys.* **2024**, *15*, 305-324.
(18) Yang, X. P.; Li, M.; Ding, Z. Y.; Li, L. Y.; Ji, C.; Wu, G. Review on Developments and Progress in Nickelate-Based Heterostructure Composites and Superconducting Thin Films. *Adv. Quantum Technol.* **2023**, *6* (1), 2200065.
(19) Zhou, X.; Qin, P.; Feng, Z.; Yan, H.; Wang, X.; Chen, H.; Meng, Z.; Liu, Z. Experimental progress on the emergent infinite-layer Ni-based superconductors. *Mater. Today* **2022**, *55*, 170-185.

(20) Osada, M.; Wang, B. Y.; Goodge, B. H.; Lee, K.; Yoon, H.; Sakuma, K.; Li, D.; Miura, M.; Kourkoutis, L. F.; Hwang, H. Y. A superconducting praseodymium nickelate with infinite layer structure. *Nano Lett.* 2020, *20*, 5735-5740.
(21) Osada, M.; Wang, B. Y.; Goodge, B. H.; Harvey, S. P.; Lee, K.; Li, D.; Kourkoutis, L. F.; Hwang, H. Y. Nickelate Superconductivity without Rare-Earth Magnetism: (La,Sr)NiO$_2$. *Adv. Mater.* 2021, *33* (45), e2104083.
(22) Zeng, S.; Li, C.; Chow, L. E.; Cao, Y.; Zhang, Z.; Tang, C. S.; Yin, X.; Lim, Z. S.; Hu, J.; Yang, P.; Ariando, A. Superconductivity in infinite-layer nickelate La$_{1-x}$Ca$_x$NiO$_2$ thin films. *Sci. Adv.* 2022, *8* (7), eabl9927.
(23) Wei, W.; Vu, D.; Zhang, Z.; Walker, F. J.; Ahn, C. H. Superconducting Nd$_{1-x}$Eu$_x$NiO$_2$ thin films using in situ synthesis. *Sci. Adv.* 2023, *9* (27), eadh3327.
(24) Pan, G. A.; Ferenc Segedin, D.; LaBollita, H.; Song, Q.; Nica, E. M.; Goodge, B. H.; Pierce, A. T.; Doyle, S.; Novakov, S.; Cordova Carrizales, D.; N'Diaye, A. T.; Shafer, P.; Paik, H.; Heron, J. T.; Mason, J. A.; Yacoby, A.; Kourkoutis, L. F.; Erten, O.; Brooks, C. M.; Botana, A. S.; Mundy, J. A. Superconductivity in a quintuple-layer square-planar nickelate. *Nat. Mater.* 2022, *21* (2), 160-164.
(25) Li, Q.; He, C.; Si, J.; Zhu, X.; Zhang, Y.; Wen, H.-H. Absence of superconductivity in bulk Nd$_{1-x}$Sr$_x$NiO$_2$. *Commun. Mater.* 2020, *1*, 16.
(26) Wang, B.-X.; Zheng, H.; Krivyakina, E.; Chmaissem, O.; Lopes, P. P.; Lynn, J. W.; Gallington, L. C.; Ren, Y.; Rosenkranz, S.; Mitchell, J. F.; Phelan, D. Synthesis and characterization of bulk Nd$_{1-x}$Sr$_x$NiO$_2$ and Nd$_{1-x}$Sr$_x$NiO$_3$. *Phys. Rev. Mater.* 2020, *4* (8), 084409.
(27) Puphal, P.; Wu, Y.-M.; Fürsich, K.; Lee, H.; Pakdaman, M.; Bruin, J. A. N.; Nuss, J.; Suyolcu, Y. E.; Aken, P. A. v.; Keimer, B.; Isobe, M.; Hepting, M. Topotactic transformation of single crystals: From perovskite to infinite-layer nickelates. *Sci. Adv.* 2021, *7* (49), eabl8091.
(28) Sun, H. L.; Huo, M. W.; Hu, X. W.; Li, J. Y.; Liu, Z. J.; Han, Y. F.; Tang, L. Y.; Mao, Z. Q.; Yang, P. T.; Wang, B. S.; Cheng, J. G.; Yao, D. X.; Zhang, G. M.; Wang, M. Signatures of superconductivity near 80 K in a nickelate under high pressure. *Nature* 2023, *621*, 493-498.
(29) Wang, M.; Wen, H.-H.; Wu, T.; Yao, D.-X.; Xiang, T. Normal and superconducting properties of La$_3$Ni$_2$O$_7$. *Chin. Phys. Lett.* 2024, *41*, 077402.
(30) Zhu, Y.; Peng, D.; Zhang, E.; Pan, B.; Chen, X.; Chen, L.; Ren, H.; Liu, F.; Hao, Y.; Li, N.; Xing, Z.; Lan, F.; Han, J.; Wang, J.; Jia, D.; Wo, H.; Gu, Y.; Gu, Y.; Ji, L.; Wang, W.; Gou, H.; Shen, Y.; Ying, T.; Chen, X.; Yang, W.; Cao, H.; Zheng, C.; Zeng, Q.; Guo, J.-g.; Zhao, J. Superconductivity in pressurized trilayer La$_4$Ni$_3$O$_{10-\delta}$ single crystals. *Nature* 2024, *631* (8021), 531-536.
(31) Zhao Dan, Y. Z., Mengwu Huo, Yu Wang, Linpeng Nie, Meng Wang, Tao Wu, Xianhui Chen. Spin-density-wave transition in double-layer nickelate La$_3$Ni$_2$O$_7$. 2024, Preprint at https://arxiv.org/abs/2402.03952.
(32) Zhang, Y.; Su, D.; Huang, Y.; Shan, Z.; Sun, H.; Huo, M.; Ye, K.; Zhang, J.; Yang, Z.; Xu, Y.; Su, Y.; Li, R.; Smidman, M.; Wang, M.; Jiao, L.; Yuan, H. High-temperature superconductivity with zero resistance and strange-metal behaviour in La$_3$Ni$_2$O$_{7-\delta}$. *Nat. Phys.* 2024, https://doi.org/10.1038/s41567-41024-02515-y.
(33) Zhang, Y.; Lin, L.-F.; Moreo, A.; Maier, T. A.; Dagotto, E. Structural phase transition, s$_\pm$-wave pairing, and magnetic stripe order in bilayered superconductor La$_3$Ni$_2$O$_7$ under pressure. *Nat. Commun.* 2024, *15* (1), 2470.
(34) Zhang, M.; Pei, C.; Wang, Q.; Zhao, Y.; Li, C.; Cao, W.; Zhu, S.; Wu, J.; Qi, Y. Effects of pressure and doping on Ruddlesden-Popper phases La$_{n+1}$Ni$_n$O$_{3n+1}$. *J. Mater. Sci. Technol.* 2024, *185*, 147-154.
(35) Yang, Q.-G.; Jiang, K.-Y.; Wang, D.; Lu, H.-Y.; Wang, Q.-H. Effective model and s$_\pm$-wave superconductivity in trilayer nickelate La$_4$Ni$_3$O$_{10}$. *Phys. Rev. B* 2024, *109* (22), L220506.
(36) Xue, J.-R.; Wang, F. Magnetism and Superconductivity in the t–J Model of La$_3$Ni$_2$O$_7$ Under Multiband Gutzwiller Approximation. *Chin. Phys. Lett.* 2024, *41* (5), 057403.
(37) Xu, M.; Huyan, S.; Wang, H.; Bud'ko, S. L.; Chen, X.; Ke, X.; Mitchell, J. F.; Canfield, P. C.; Li, J.; Xie, W. Pressure-Dependent "Insulator–Metal–Insulator" Behavior in Sr-Doped La$_3$Ni$_2$O$_7$. *Adv. Electron. Mater.* 2024, 2400078.
(38) Wú, W.; Luo, Z.; Yao, D.-X.; Wang, M. Superexchange and charge transfer in the nickelate superconductor La$_3$Ni$_2$O$_7$ under pressure. *Sci. China-Phys. Mech. Astron.* 2024, *67* (11), 117402.
(39) Wang, L.; Li, Y.; Xie, S.-Y.; Liu, F.; Sun, H.; Huang, C.; Gao, Y.; Nakagawa, T.; Fu, B.; Dong, B.; Cao, Z.; Yu, R.; Kawaguchi, S. I.; Kadobayashi, H.; Wang, M.; Jin, C.; Mao, H.-k.; Liu, H. Structure Responsible for the Superconducting State in La$_3$Ni$_2$O$_7$ at High-Pressure and Low-Temperature Conditions. *J. Am. Chem. Soc.* 2024, *146* (11), 7506-7514.
(40) Wang, J.-X.; Ouyang, Z.; He, R.-Q.; Lu, Z.-Y. Non-Fermi liquid and Hund correlation in La$_4$Ni$_3$O$_{10}$ under high pressure. *Phys. Rev. B* 2024, *109* (16), 165140.
(41) Wang, H. Z.; Chen, L.; Rutherford, A.; Zhou, H. D.; Xie, W. W. Long-Range Structural Order in a Hidden Phase of Ruddlesden-Popper Bilayer Nickelate La$_3$Ni$_2$O$_7$. *Inorg. Chem.* 2024, *63* (11), 5020-5026.
(42) Wang, G.; Wang, N. N.; Shen, X. L.; Hou, J.; Ma, L.; Shi, L. F.; Ren, Z. A.; Gu, Y. D.; Ma, H. M.; Yang, P. T.; Liu, Z. Y.; Guo, H. Z.; Sun, J. P.; Zhang, G. M.; Calder, S.; Yan, J. Q.; Wang, B. S.; Uwatoko, Y.; Cheng, J. G. Pressure-Induced Superconductivity In Polycrystalline La$_3$Ni$_2$O$_{7-\delta}$. *Phys. Rev. X* 2024, *14* (1), 011040.
(43) Tian, Y.-H.; Chen, Y.; Wang, J.-M.; He, R.-Q.; Lu, Z.-Y. Correlation effects and concomitant two-orbital s$_\pm$-wave superconductivity in La$_3$Ni$_2$O$_7$ under high pressure. *Phys. Rev. B* 2024, *109* (16), 165154.
(44) Sui, X.; Han, X.; Jin, H.; Chen, X.; Qiao, L.; Shao, X.; Huang, B. Electronic properties of the bilayer nickelates R$_3$Ni$_2$O$_7$ with oxygen vacancies (R=La or Ce). *Phys. Rev. B* 2024, *109* (20), 205156.
(45) Siqi Wu, Z. Y., Xin Ma, Jianhui Dai, Ming Shi, Hui-Qiu Yuan, Hai-Qing Lin, Chao Cao. Ac$_3$Ni$_2$O$_7$ and La$_2$AeNi$_2$O$_6$F (Ae = Sr, Ba): Benchmark Materials for Bilayer Nickelate Superconductivity. 2024, Preprint at https://arxiv.org/abs/2403.11713.
(46) Sebastien N. Abadi, K.-J. X., Eder G. Lomeli, Pascal Puphal, Masahiko Isobe, Yong Zhong, Alexei V. Fedorov, Sung-Kwan Mo, Makoto Hashimoto, Dong-Hui Lu, Brian Moritz, Bernhard Keimer, Thomas P. Devereaux, Matthias Hepting, Zhi-Xun Shen. Electronic structure of the alternating monolayer-trilayer phase of La$_3$Ni$_2$O$_7$. 2024, Preprint at https://arxiv.org/abs/2402.07143.
(47) Sakakibara, H.; Kitamine, N.; Ochi, M.; Kuroki, K. Possible High T$_c$ Superconductivity in La$_3$Ni$_2$O$_7$ under High Pressure through Manifestation of a Nearly Half-Filled Bilayer Hubbard Model. *Phys. Rev. Lett.* 2024, *132* (10), 106002.
(48) Qu, X.-Z.; Qu, D.-W.; Chen, J.; Wu, C.; Yang, F.; Li, W.; Su, G. Bilayer t-J-J$_\perp$ Model and Magnetically Mediated Pairing in the Pressurized Nickelate La$_3$Ni$_2$O$_7$. *Phys. Rev. Lett.* 2024, *132* (3), 036502.
(49) Ningning Wang, G. W., Xiaoling Shen, Jun Hou, Jun Luo, Xiaoping Ma, Huaixin Yang, Lifen Shi, Jie Dou, Jie Feng, Jie Yang, Yunqing Shi, Zhian Ren, Hanming Ma, Pengtao Yang, Ziyi Liu, Yue Liu, Hua Zhang, Xiaoli Dong, Yuxin Wang, Kun Jiang, Jiangping Hu, Stuart Calder, Jiaqiang Yan, Jianping Sun, Bosen Wang, Rui Zhou, Yoshiya Uwatoko, Jinguang Cheng. Bulk high-temperature superconductivity in the high-pressure tetragonal phase of bilayer La$_2$PrNi$_2$O$_7$. 2024, Preprint at https://arxiv.org/abs/2407.05681.
(50) Lu, C.; Pan, Z.; Yang, F.; Wu, C. Interlayer-Coupling-Driven High-Temperature Superconductivity in La$_3$Ni$_2$O$_7$ under Pressure. *Phys. Rev. Lett.* 2024, *132* (14), 146002.
(51) Li, Q.; Zhang, Y.-J.; Xiang, Z.-N.; Zhang, Y.; Zhu, X.; Wen, H.-H. Signature of superconductivity in pressurized La$_4$Ni$_3$O$_{10}$. *Chin. Phys. Lett.* 2024, *41*, 017401.
(52) Li, F.; Wang, S.; Ma, C.; Wang, X.; Liu, C.; Fan, C.; Han, L.; Wang, S.; Tao, X.; Zhang, J. Flux Growth of Trilayer La$_4$Ni$_3$O$_{10}$ Single Crystals at Ambient Pressure. *Cryst. Growth Des.* 2024, *24* (1), 347-354.
(53) LaBollita, H.; Kapeghian, J.; Norman, M. R.; Botana, A. S. Electronic structure and magnetic tendencies of trilayer La$_4$Ni$_3$O$_{10}$ under pressure: Structural transition, molecular orbitals, and layer differentiation. *Phys. Rev. B* 2024, *109* (19), 195151.
(54) Jiang, R.; Hou, J.; Fan, Z.; Lang, Z.-J.; Ku, W. Pressure Driven Fractionalization of Ionic Spins Results in Cupratelike High-T$_c$ Superconductivity in La$_3$Ni$_2$O$_7$. *Phys. Rev. Lett.* 2024, *132* (12), 126503.

(55) Geisler, B.; Hamlin, J. J.; Stewart, G. R.; Hennig, R. G.; Hirschfeld, P. J. Structural transitions, octahedral rotations, and electronic properties of $A_3Ni_2O_7$ rare-earth nickelates under high pressure. *npj Quantum Mater.* **2024**, *9* (1), 38.

(56) Frank Lechermann, S. B., Ilya M. Eremin. Electronic instability, layer selectivity and Fermi arcs in $La_3Ni_2O_7$. **2024**, Preprint at https://arxiv.org/abs/2403.12831.

(57) Dong, Z.; Huo, M.; Li, J.; Li, J.; Li, P.; Sun, H.; Gu, L.; Lu, Y.; Wang, M.; Wang, Y.; Chen, Z. Visualization of oxygen vacancies and self-doped ligand holes in $La_3Ni_2O_{7-\delta}$. *Nature* **2024**, *630*, 847-852.

(58) Cui, T.; Choi, S.; Lin, T.; Liu, C.; Wang, G.; Wang, N.; Chen, S.; Hong, H.; Rong, D.; Wang, Q.; Jin, Q.; Wang, J.-O.; Gu, L.; Ge, C.; Wang, C.; Cheng, J.-G.; Zhang, Q.; Si, L.; Jin, K.-j.; Guo, E.-J. Strain-mediated phase crossover in Ruddlesden–Popper nickelates. *Commun. Mater.* **2024**, *5* (1), 32.

(59) Craco, L.; Leoni, S. Strange metal and coherence-incoherence crossover in pressurized $La_3Ni_2O_7$. *Phys. Rev. B* **2024**, *109* (16), 165116.

(60) Chen, X.; Zhang, J.; Thind, A. S.; Sharma, S.; LaBollita, H.; Peterson, G.; Zheng, H.; Phelan, D. P.; Botana, A. S.; Klie, R. F.; Mitchell, J. F. Polymorphism in the Ruddlesden–Popper Nickelate $La_3Ni_2O_7$: Discovery of a Hidden Phase with Distinctive Layer Stacking. *J. Am. Chem. Soc.* **2024**, *146* (6), 3640-3645.

(61) Yazhou Zhou, J. G., Shu Cai, Hualei Sun, Pengyu Wang, Jinyu Zhao, Jinyu Han, Xintian Chen, Qi Wu, Yang Ding, Meng Wang, Tao Xiang, Ho-kwang Mao, Liling Sun. Evidence of filamentary superconductivity in pressurized $La_3Ni_2O_7$ single crystals. **2023**, Preprint at https://arxiv.org/abs/2311.12361.

(62) Pascal Puphal, P. R., Niklas Enderlein, Yu-Mi Wu, Giniyat Khaliullin, Vignesh Sundaramurthy, Tim Priessnitz, Manuel Knauft, Lea Richter, Masahiko Isobe, Peter A. van Aken, Hidenori Takagi, Bernhard Keimer, Y. Eren Suyolcu, Björn Wehinger, Philipp Hansmann, Matthias Hepting. Unconventional crystal structure of the high-pressure superconductor $La_3Ni_2O_7$. **2023**, Preprint at https://arxiv.org/abs/2312.07341.

(63) Mingxin Zhang, C. P., Xian Du, Yantao Cao, Qi Wang, Juefei Wu, Yidian Li, Yi Zhao, Changhua Li, Weizheng Cao, Shihao Zhu, Qing Zhang, Na Yu, Peihong Cheng, Jinkui Zhao, Yulin Chen, Hanjie Guo, Lexian Yang, Yanpeng Qi. Superconductivity in trilayer nickelate $La_4Ni_3O_{10}$ under pressure. **2023**, Preprint at https://arxiv.org/abs/2311.07423.

(64) Luo, Z.; Hu, X.; Wang, M.; Wú, W.; Yao, D.-X. Bilayer Two-Orbital Model of $La_3Ni_2O_7$ under Pressure. *Phys. Rev. Lett.* **2023**, *131* (12), 126001.

(65) LIU, Z.; LI, Q.; ZHOU, X.; HAO, J.; DAI, Y.; WEN, H.-H. Infrared spectroscopic study of $Nd_4Ni_3O_{10}$. *Sci Sin-Phys Mech Astron,* **2023**, *53* (12), 127416.

(66) Liu, Y.-B.; Mei, J.-W.; Ye, F.; Chen, W.-Q.; Yang, F. s±-Wave Pairing and the Destructive Role of Apical-Oxygen Deficiencies in $La_3Ni_2O_7$ under Pressure. *Phys. Rev. Lett.* **2023**, *131* (23), 236002.

(67) Yuecong Liu, M. O., Haifeng Chu, Huan Yang, Qing Li, Yingjie Zhang, Hai-Hu Wen. Growth and characterization of the $La_3Ni_2O_{7-\delta}$ thin films: dominant contribution of the $d_{x^2-y^2}$ orbital at ambient pressure. **2024**, Preprint at https://arxiv.org/abs/2406.08789.

(68) Zhang, J.; Zheng, H.; Chen, Y.-S.; Ren, Y.; Yonemura, M.; Huq, A.; Mitchell, J. F. High oxygen pressure floating zone growth and crystal structure of the metallic nickelates $R_4Ni_3O_{10}$ (R=La, Pr). *Phys. Rev. Mater.* **2020**, *4* (8), 083402.

(69) Hongquan Liu, C. X., Shengjie Zhou, Hanghui Chen. Role of crystal-field-splitting and longe-range-hoppings on superconducting pairing symmetry of $La_3Ni_2O_7$. **2023**, Preprint at https://arxiv.org/abs/2311.07316.

(70) Yang Shen, M. Q., Guang-Ming Zhang. Effective Bi-Layer Model Hamiltonian and Density-Matrix Renormalization Group Study for the High-$T_c$ Superconductivity in $La_3Ni_2O_7$ under High Pressure. *Chin. Phys. Lett.* **2023**, *40*, 127401.

(71) Yang, Y.-f.; Zhang, G.-M.; Zhang, F.-C. Interlayer valence bonds and two-component theory for high-$T_c$ superconductivity of $La_3Ni_2O_7$ under pressure. *Phys. Rev. B* **2023**, *108* (20), L201108.

(72) Sakakibara, H.; Ochi, M.; Nagata, H.; Ueki, Y.; Sakurai, H.; Matsumoto, R.; Terashima, K.; Hirose, K.; Ohta, H.; Kato, M.; Takano, Y.; Kuroki, K. Theoretical analysis on the possibility of superconductivity in the trilayer Ruddlesden-Popper nickelate $La_4Ni_3O_{10}$ under pressure and its experimental examination: Comparison with $La_3Ni_2O_7$. *Phys. Rev. B* **2024**, *109* (14), 144511.

(73) Li, F.; Guo, N.; Zheng, Q.; Shen, Y.; Wang, S.; Cui, Q.; Liu, C.; Wang, S.; Tao, X.; Zhang, G.-M.; Zhang, J. Design and synthesis of three-dimensional hybrid Ruddlesden-Popper nickelate single crystals. *Phys. Rev. Mater.* **2024**, *8* (5), 053401.

(74) Woollam, J. A.; Somoano, R. B.; O'Connor, P. Positive Curvature of the $H_{c2}$-versus-$T_c$ Boundaries in Layered Superconductors. *Phys. Rev. Lett.* **1974**, *32* (13), 712-714.

(75) Landau, L. D.; Ginzburg, V. L. On the theory of superconductivity. *Zh. Eksp. Teor. Fiz.* **1950**, *20*, 1064-1082.

(76) Li, J.; Chen, C.-Q.; Huang, C.; Han, Y.; Huo, M.; Huang, X.; Ma, P.; Qiu, Z.; Chen, J.; Hu, X.; Chen, L.; Xie, T.; Shen, B.; Sun, H.; Yao, D.-X.; Wang, M. Structural transition, electric transport, and electronic structures in the compressed trilayer nickelate $La_4Ni_3O_{10}$. *Sci. China-Phys. Mech. Astron.* **2024**, *67* (11), 117403.